\documentclass[aps,prl,reprint,groupedaddress,nofootinbib]{revtex4-2}
\usepackage{mathtools,slashed}
\usepackage{graphicx}
\usepackage{dcolumn}
\usepackage{bm}
\usepackage{amsmath,amsfonts,amssymb,mathtools}
\usepackage{graphicx}
\usepackage{enumerate}
\usepackage{color}
\usepackage{booktabs}
\usepackage{changepage}
\usepackage{hyperref}
\usepackage[normalem]{ulem}
\def\edotb{\vec{E}.\vec{B}}
\def\Mp{M_{\rm Pl}} 
\begin{document}

\title{CMB constraints on U(1) Axion warm inflation}
\author{Asma Alaei$^a$}
\email{asmaalaei6999@gmail.com}
\author{Sukannya Bhattacharya$^b$}
\email{sukannya.bhattacharya@ift.csic.es}
\author{Vahid Kamali$^{acde}$}
\email{vahid.kamali2@mcgill.ca}
\affiliation{${}^a$Department of Physics, Bu-Ali Sina (Avicenna) University, Hamedan 65178, 016016, Iran\\
${}^b$ Instituto de F\'isica T\'eorica UAM/CSIC, Calle Nicol\'as Cabrera 13-15, Cantoblanco, 28049, Madrid, Spain\\
${}^c$Department of Physics, McGill University, Montr\'{e}al, QC, H3A 2T8, Canada\\
${}^d$School of Continuing Studies, McGill University, Montr\'{e}al, QC, H3A 2T5, Canada\\
${}^e$Trottier Space Institute, Department of Physics, McGill University, Montr\'{e}al, QC, H3A 2T8, Canada
}

\date{\today}

\begin{abstract}
 In this work, we propose a model of warm inflation driven by axion-like particles interacting with $U(1)$ gauge fields, with implications for the early universe's thermal evolution.
  By extending traditional warm inflation models, we introduce a dissipation mechanism through the thermal fluctuations of electromagnetic fields, leading to a non-trivial backreaction on the inflaton's dynamics. Our results are consistent with CMB observations, even for a natural sub-Plankian Axion decay constant $f\leq M_p$. We present precise constraints on the model’s free parameters, using CAMB and CosmoMC codes. These findings offer new insights into the thermal history of the universe and the nature of inflationary dynamics.
\end{abstract}


\maketitle

\section{Introduction}
While the standard inflation model provides an elegant framework for the early time evolution of the universe \cite{Starobinsky:1980te,Guth:1980zm,Albrecht:1982wi,Linde:1981mu}, certain aspects of the model building, as well as the details of the reheating mechanism following inflation remain poorly understood despite continued theoretical efforts~\cite{Kofman:1994rk, Shtanov:1994ce,Bassett:2005xm,Frolov:2010sz,Allahverdi:2010xz,Amin:2014eta,Lozanov:2019jxc,Allahverdi:2020bys}.

As a modification of the standard cold inflation paradigm, the motivation for warm inflation is to realise inflation in a more general setting where the {\it{inflaton}} field co-exists with radiation, while the energy stored in the inflaton field continuously dissipates into radiation \cite{Berera:1995ie, Bastero-Gil:2016qru, Kamali:2023lzq}. This dissipation can occur through interactions with other fields, leading to a thermal bath that modifies the background and perturbations of inflation dynamics. Warm inflation is thus a natural extension of the standard inflationary paradigm, where the inflaton field not only drives accelerated expansion but also efficiently transfers energy to radiation, maintaining a non-zero temperature throughout the inflationary phase, hence allowing for a smooth transition from inflation to reheating. The modifications to the power spectrum of primordial perturbations in the warm setup can also be constrained with the CMB data\cite{a38, Aghanim:2018eyx, Aghanim:2019ame,Motaharfar:2018zyb,Kamali:2019ppi,Berghaus:2019whh,Visinelli:2016rhn,Benetti:2016jhf,Videla:2017asf,Bastero-Gil:2017wwl,Arya:2017zlb,Gomes:2018uhv,Bastero-Gil:2018uep,Das:2018rpg,Bastero-Gil:2019rsp,Montefalcone:2022owy,Montefalcone:2022jfw,Das:2022ubr,Mukuno:2024yoa,Kumar:2024hju}.

The interaction of the inflaton with the fields in the thermal bath to realise such dissipation is, however, a crucial part of the model building to preserve the flatness of the potential. Light mediators, e.g. fermions, can introduce large thermal corrections to the inflaton potential, ruining slow-roll; however such models can be realised with additional symmetries~\cite{Bastero-Gil:2016qru}. Heavy mediators preserve the flatness of the inflaton potential, however, can only lead to sizeable dissipation when present in large quantities. Recently, warm inflation models have gained significant attention in the context of axion-like particles, which arise in several high-energy theories, including string theory and various models of particle physics beyond the Standard Model \cite{Kamali:2019ppi}\cite{Berghaus:2019whh, DeRocco:2021rzv,Laine:2021ego, Montefalcone:2022jfw, Mukuno:2024yoa,Berghaus:2025dqi,ORamos:2025uqs,Broadberry:2025ggb,Bhattacharya:2025guc}. Axions, which are pseudoscalar particles, interacting with (non-)Abelian gauge fields, could provide a thermal radiation bath during inflation. It has been explored in previous studies \cite{Ferreira:2017lnd,Visinelli:2011jy}. 

In this paper, we propose a model of warm inflation in which the interaction between axion inflaton and $U(1)$ gauge fields generates a thermal bath during inflation, and a sub-Planckian decay constant is able to achieve enough number of e-folds and observationally compatible model\footnote{Similar scenarios of warm inflation has been realised in previous literature with cosine potential where inflaton is a pNGB coupled to another pNGB field~\cite{Mishra:2011vh} and in the context of SU(N) gauge fields coupled to axions~\citep{Kamali:2024qme,Mukuno:2024yoa,Berghaus:2019whh,DeRocco:2021rzv}. See~\cite{Bastero-Gil:2016qru} for a concrete model of \textit{warm little inflation} by directly coupling the inflaton (pNGB) to a few light fermions. References~\cite{Ferreira:2017wlx} and~\cite{Bhattacharya:2025guc} derived limits for thermalization in axion inflation coupled to $U(1)$ and $ SU(2)$ gauge fields respectively. }.
In particular, here, the plasma consists of $U(1)$ gauge fields, which is continuously replenished by axion decay and we investigate the {\it{thermal backreaction}} of the plasma on the evolution of the inflaton. Our setup also includes fermions charged under the $U(1)$ gauge field, which are crucial for achieving a thermalized plasma and analyzing its backreaction on the inflaton dynamics. \cite{Hassan:2024bvb,Fujita:2025zoa}. 

By representing the Chern-Simons term with a thermal average of the local fluctuations of the electromagnetic fields, we utilize the linear response method and real-time thermal field theory~\cite{Bellac,McLerran:1990de,Khlebnikov:1992bx} to quantify the dissipation. We evaluate the required spectral densities using the HTL effective theory of the plasma \cite{LeBellac,Khlebnikov:1992bx}.

This approach allows us to model the thermal influence on the background of the system and evaluate a power spectrum for the dominant thermal fluctuations. This is an interesting approach, since the gauge field fluctuations reach large amplitudes within an instability band closer to horizon exit~\cite{Anber:2009ua,Barnaby:2011vw,Barnaby:2010vf,Cook:2011hg,Ozsoy:2014sba,Cheng:2015oqa,Peloso:2016gqs,Gorbar:2021rlt,Durrer:2024ibi,Corba:2024tfz,Jimenez:2017cdr}, and limits on back-reaction of the gauge fields on the axion are typically studied in the context of cold inflation paradigm~\cite{Domcke:2020zez,Peloso:2022ovc,Garcia-Bellido:2023ser,vonEckardstein:2023gwk,Alam:2024fid,He:2025ieo,Figueroa:2023oxc,Sharma:2024nfu,Caravano:2021bfn}. Our approach assumes thermalization, in fact thermalised initial conditions, and the thermal bath prevails exponential expansion.

To check the consistency with data, we constrain our model with the latest CMB data (Planck 2018) using MCMC simulations\footnote{We use CAMB (Code for Anisotropies in the Microwave Background) and CosmoMC (Cosmological Monte Carlo) codes, which are widely used to calculate the CMB power spectrum and constrain cosmological parameters.} to find the best fit values of the model parameters. These tools allow us to derive precise constraints on key parameters, such as the axion decay constant $f$, the axion mass $m_{\phi}$, and the dissipation parameter $Q$. Interestingly, $Q$ arises from the thermal backreactions, and therefore it is directly linked to the Chern-Simons term. It is worth noting here that we assume the backreaction to be large enough so that thermalization is be possible, but not strong enough to ruin slow-roll, at least at CMB scales. Our results provide new insights into the possibility of thermal effects present during inflation and the potential signatures of warm inflationary models in the CMB. 

In Section~\ref{sec:themodel}, we present our setup and considering thermal treatment of the backreaction, reach an expression for the dissipation coefficient. In Section~\ref{sec:wi}, the system is presented in the warm paradigm with the background evolution and warm power spectrum. In Section~\ref{sec:analysis}, we confront our setup with the latest CMB data and constrain the model parameters. In Section~\ref{sec:condis} we conclude our analysis and discuss our results for the parameters in view of existing bounds in the literature. 

\section{The model}
\label{sec:themodel}
The Lagrangian of the model contains axion$~\phi$ as the inflaton and fermions$~\psi$ charged under U(1) gauge-fields $A_{\mu}$ and  a Chern-Simons coupling between inflaton and gauged fields in a curved space-time:
\begin{eqnarray}\label{eq:Lagrangian}
     \mathcal{L}= \sqrt{-g}[-\frac{R}{2}-\frac{1}{2}F_{\mu\nu}F^{\mu\nu}-\frac{1}{2}(\partial_{\mu}\phi)^2-V(\phi)\\
     \nonumber
  +\bar{\psi}(i \slashed{D} -m_f)\psi -\frac{\lambda}{4\pi f}\phi F_{\mu\nu}\tilde{F}^{\mu\nu}]
\end{eqnarray}
where $F_{\mu \nu} \equiv \partial_\mu A_\nu - \partial_\nu A_\mu$ is the U(1) field strength, while  $\tilde{F}^{\mu \nu} \equiv \frac{1}{2} \frac{\epsilon^{\mu \nu \alpha \beta}}{\sqrt{-g}}F_{\alpha \beta}$ is its dual, with $\epsilon^{\mu \nu \alpha \beta}$ totally anti-symmetric (and $\epsilon^{0123}=+1$). $g$ the determinant of the underlying FLRW metric with the line element $ds^2=-dt^2+a^2(t)d\vec{x}^2$, $a(t)$ being the scale factor. 
\begin{table}[t]
\caption{Particle content and $U(1)$ charge assignment of the effective model.}
\label{tab:fieldcontent}
\begin{ruledtabular}
\begin{tabular}{l c c c}
\textbf{Field} & \textbf{Spin} & \textbf{$U(1)$ charge} & \textbf{Role in the model} \\
\hline
$A_\mu$ & $1$ & Gauge connection & \parbox[t]{0.55\columnwidth}{Abelian gauge field forming the thermal radiation sector; $F_{\mu\nu}$ and $F_{\mu\nu}\tilde F^{\mu\nu}$ are gauge-invariant operators.} \\
\hline
$\phi$ & $0$ & $0$ & \parbox[t]{0.55\columnwidth}{Axion-like inflaton and gauge singlet; couples via the shift-symmetric Chern--Simons operator $\phi F_{\mu\nu}\tilde F^{\mu\nu}/f$, not via a gauge charge.} \\
\hline
$\psi$ & $1/2$ & $+e$ (Dirac field) & \parbox[t]{0.55\columnwidth}{Charged vectorlike fermion in the plasma; antiparticle has charge $-e$, with covariant derivative $D_\mu = \nabla_\mu + i e A_\mu$.} \\
\end{tabular}
\end{ruledtabular}
\end{table}
Table~\ref{tab:fieldcontent} makes explicit that the inflaton is not charged under the $U(1)$ gauge symmetry. Therefore the last term in Eq.~\eqref{eq:Lagrangian} does not break the gauge symmetry: $F_{\mu\nu}\tilde F^{\mu\nu}$ is gauge invariant, while the axion-like field $\phi$ is a neutral pseudoscalar. The charged matter sector is carried by the vectorlike fermion $\psi$, which is included to provide the finite-temperature charged plasma response used below. For $N_f$ identical charged fermions, the replacement $m_D^2=e^2T^2/3\rightarrow e^2N_fT^2/3$ should be made in the Debye mass; in the phenomenological analysis we use one effective charged species.

We consider the following periodic potential for the scalar field
 \begin{eqnarray} \label{eq:Potential}
     V(\phi)=m_{\phi}^2f^2\bigg(1+\cos\bigg(\frac{\phi}{f}\bigg)\bigg),
 \end{eqnarray}
 where the shift symmetry protects it from thermal corrections.

Since the inflaton couples through the Chern–Simons operator $\phi F\tilde F$, which respects a perturbative shift symmetry, perturbative finite-temperature loops from the $U(1)$ plasma do not generate a thermal mass term $\,\propto T^2\phi^2$ for the inflaton. 
Instead, the thermal bath modifies the \emph{dissipative} and \emph{kinetic} sectors, encoded here by the HTL-controlled $\Upsilon(T)$ and, subdominantly, by higher-derivative operators such as $c(T)(\partial\phi)^2$. 
In our adiabatic, strong-interaction window these effects are captured by $\Upsilon(T)$, while kinetic renormalizations remain subleading.
 $m_{\phi}$ is axion mass and $f$ is decay constant. We will study our model in natural units, $\hbar
 =c= \kappa_b=1$.  
 
 Variation of the above Lagrangian (\ref{eq:Lagrangian}) with respect to the scalar field leads to the following modified Klein-Gordon equation for the axion background
 \begin{eqnarray}\label{eq:Backgroung E.O.M}
     \ddot{\phi}+3 H\dot{\phi}+V_{\phi}(\phi)=\frac{\lambda}{2\pi f}(\edotb), 
     \end{eqnarray}
where $V_{\phi}(\phi)=\frac{dV}{d\phi}$ and the source term on the right-hand side of Eq.(\ref{eq:Backgroung E.O.M}) can produce a thermal bath through dissipation. This system has two important pictures of the thermal process according to their physical origin. The first one includes the scattering between axion and gauge fields that normally leads to a weak dissipative regime of warm inflation where the backreaction term, i.e the R.H.S of Eq.(\ref{eq:Backgroung E.O.M}), is negligible at the background level \cite{Wu:2023lcz}. The second one, which we are interested in this work, involves a collective dissipative effect of the thermal bath of gauge fields.
    
To study the thermal backreaction of the plasma on the evolution of the axion, we replace the Chern-Simons term $\vec{E}.\vec{B}$ in Eq.(\ref{eq:Backgroung E.O.M}) by the thermal average of the local fluctuations of the electromagnetic fields, $\langle \edotb \rangle$, and utilize the linear response method in the context of real-time thermal field theory \cite{Bellac} (see Appendix~\ref{APP1}). 
\begin{equation} \label{eq:thermal interaction}
     \frac{\lambda }{2\pi f}\langle \edotb \rangle=\frac{\lambda}{2\pi f}\langle \edotb \rangle_0+\int d^4x' \Sigma _R(x-x') \delta \phi (t').
     \end{equation}
Here,  $\Sigma _R(x-x')$ is a retarded correlation of the fields, containing non-local information, which depends on the interaction Lagrangian $\mathcal{L}_{\rm int}$, related to the retarded Green's function $\Sigma_{\rho}(x-x')$ as:
      \begin{eqnarray}\label{eq:sigmaR}
       \Sigma _R(x-x') &=& \Theta (t-t') \Sigma_{\rho} (x-x')\\ 
       \Sigma_{\rho} (x-x') &=& i \bigg(\frac{\partial f(\phi)}{\partial \phi}\bigg)^2 \langle [g(X(x)), g(X(x'))] \rangle ,\\ 
       \rm{where,} \nonumber \\
       \mathcal{L}_{\rm int} &=& f(\phi)g(X), \nonumber
       \end{eqnarray}
$X$ being the radiation component.

In case of the Lagrangian in Eq.~\ref{eq:Lagrangian}, therefore,   
\begin{equation}\label{eq:Greens}
 \Sigma_{\rho} (x-x') = -i \frac{\lambda ^2}{4\pi ^2 f^2} \langle [(\edotb)(x), (\edotb)(x')]\rangle . 
 \end{equation}
 At the level of no interaction specified by zero index, the CP conservation vanishes the first term in eq.(\ref{eq:thermal interaction}), so that,
     \begin{eqnarray} \label{eq:thermalint}
    \frac{\lambda }{2\pi f}\langle \edotb \rangle &=&-i \frac{\lambda ^2}{4\pi ^2 f^2} \int dt' \Theta(t-t') \\  \nonumber
    &\times& \int d^3x'\langle [(\edotb)(x), (\edotb)(x')]\rangle .
     \end{eqnarray}
In the adiabatic approximation, when the inflaton is varying on a much larger timescale than the response (or relaxation) time of dissipation, i.e., $\frac{\dot{\phi}}{\phi} \ll \tau ^{-1}$, then the source term of Eq.~\ref{eq:Backgroung E.O.M}, i.e. $ \frac{\lambda }{2\pi f}\langle \edotb \rangle $ can be written in the form of a friction term $\Upsilon \dot{\phi}$ in the background equation: 
\begin{eqnarray}\label{eq:warmbackphi}
   \ddot{\phi}+(3H+\Upsilon)\dot{\phi}+V_{\phi}(\phi)=0.
\end{eqnarray}
The dissipation parameter $\Upsilon $ is in general a function of $\phi$ and temperature $T$, and is responsible for transferring the kinetic energy of inflaton (axion) to the radiation (gauge field) sector. It is possible to find the form of $\Upsilon$ from the retarded green function $\Sigma_{\rho} (x-x')$ \cite{Bastero-Gil:2010dgy}. 
\begin{eqnarray}\label{eq:upsilon}
\Upsilon &=& i\lim_{\omega\rightarrow~0}\frac{\partial\tilde{\Sigma}_{\rho}(\omega,0)}{\partial\omega},\\
{\rm where,} \\ \nonumber
\tilde{\Sigma}_{\rho}(\omega,\vec{k})&=& \int d^4y  \Sigma_{\rho} (t,\vec{y}) e ^{-i\omega t}e ^{i\vec{k}.\vec{y}}. \nonumber
\end{eqnarray}
The retarded Green function $\Sigma_{\rho} (x-x')$ contains all the information related to back-reaction in the thermal setting, which can be evaluated for our case in terms of $\beta = 1/T $ as (see Appendix~\ref{APP1})
\begin{eqnarray}\label{eq:Greenstemp}
\nonumber
    \tilde{\Sigma}_{\rho}(\omega,\vec{k}) &=& -i \frac{\lambda ^2}{4\pi ^2 f^2} (1-e^{-\beta\omega})\\ \nonumber 
    & \times & \epsilon^{\alpha\beta\gamma\tau}\epsilon^{\mu\nu\rho\sigma}\langle\partial_{\alpha}A_{\beta}\partial_{\gamma}A_{\tau}\partial'_{\mu}A_{\nu}\partial'_{\rho}A_{\sigma}\rangle\\
    &=& -i\frac{\lambda^2\omega}{768 \pi^3f^2}\bigg(\omega^2-m_D^2\bigg)^{1/2}\bigg(\omega^2+2m_D^2\bigg)\nonumber \\
    &\times& (1-e^{-\beta\omega})\bigg[\frac{1}{e^{\beta \omega/2}-1}+\frac{1}{(e^{\beta \omega/2}-1)^2}\bigg].
\end{eqnarray}
Here, $m_D$ is the thermal mass, that can be written in terms of the Debye length~\cite{Hassan:2024bvb} $m_D\sim \lambda _D^{-1}\sim eT$, where $e$ is the fermionic charge. { The fermionic mass doesn't appear in this expression since the plasma temperature is substantially greater than the fermionic mass so that the thermal bath is relativistic.} To find the above form of the correlation function, we have used the tree-level calculation in thermal field theory \cite{Bellac}. Using the expression in~\eqref{eq:upsilon},
\begin{equation}\label{eq:upsilonexpression}
\Upsilon = \frac{e^3}{192\pi^3}\frac{\lambda ^2}{f^2}T^3.
\end{equation}

If the charged fermions are removed, $N_f=0$ and hence $m_D^2=e^2N_fT^2/3=0$. Repeating the Kubo calculation in this limit gives a retarded spectral function with no term linear in $\omega$ at $\omega\rightarrow0$; equivalently,
\begin{equation}
\Upsilon_{N_f=0}=\frac{\lambda^2}{(4\pi f)^2}\lim_{\omega\rightarrow0}\frac{1}{\omega}\rho^{\rm ret}_{E\!\cdot\!B\,E\!\cdot\!B}(\omega,\mathbf{k}=0)\bigg|_{m_D=0}=0.
\label{eq:upsilonzero_main}
\end{equation}
Thus the theory without charged fermions is not gauge-inconsistent, but it does not realize the warm-inflation mechanism studied here: the collective Abelian plasma response vanishes, $Q=\Upsilon/(3H)=0$, and the system reduces to the usual cold axion-$U(1)$ setup with gauge-field production rather than local thermal friction. We show the small-$\omega$ limit explicitly in Appendix~\ref{APPfermionZero}.

 Therefore, the dissipation parameter obtained here is cubic in temperature. Correspondingly, 
\begin{equation}
Q\equiv \frac{\Upsilon}{3H} = \frac{e^3}{576\pi^3}\frac{\lambda ^2}{f^2}\frac{T^3}{H}
\label{eq:Q1}
\end{equation}
We will show in the next section that in the adiabatic limit, we can consider $Q$ to be nearly constant\footnote{In an ultrarelativistic $U(1)$ plasma with $N_f$ charged fermions, the Debye mass and electric conductivity are
\begin{equation}
m_D^2 \;=\; \frac{e^2 N_f T^2}{3}\,, 
\qquad 
\sigma_E \;\simeq\; \frac{c_\sigma\,T}{e^2\ln(1/e)}\,, 
\end{equation}
with $c_\sigma=\mathcal O(1)$. Evaluating the Kubo formula for the dissipation coefficient,
\begin{equation}
\Upsilon \;=\; \frac{\lambda^2}{(4\pi f)^2}\,\lim_{\omega\to0}\,\frac{1}{\omega}\,
\rho_{E\!\cdot\!B\,E\!\cdot\!B}^{\,\rm (ret)}(\omega,\mathbf{k}=0)\,,
\end{equation}
using HTL-resummed photon propagators gives a finite $\omega\!\to\!0$ limit dominated by spacelike Landau damping at momenta $k\sim m_D$. Two regimes emerge:
\begin{align}
\text{diffusive:}\quad & \omega \ll \frac{m_D^2}{\sigma_E}\!:\quad 
\Upsilon \;\simeq\; C_{\rm diff}\,\frac{\lambda^2}{f^2}\,\frac{m_D^4}{\sigma_E}\,T\\ 
\;\sim\; C_{\rm diff}\,\frac{\lambda^2}{f^2}\,e^{6}\ln\!\frac{1}{e}\;T^3,\\[4pt]
\text{collisionless:}\quad & \frac{m_D^2}{\sigma_E}\ll \omega \ll m_D\!:\quad 
\Upsilon \;\simeq\; C_{\rm coll}\,\frac{\lambda^2}{f^2}\,m_D^{2}\,T^{2}\\
\;\sim\; C_{\rm coll}\,\frac{\lambda^2}{f^2}\,e^{6}\,T^3,
\end{align}
with $C_{\rm diff},C_{\rm coll}=\mathcal O(10^{-2}\!\!-\!1)$. Using $\lambda=q\,\alpha$ with $\alpha=e^2/(4\pi)$ recovers the overall $\Upsilon\propto e^{7}T^3/f^2$ scaling quoted below, up to a mild $\ln(1/e)$ in the strict $\omega\!\to\!0$ limit. This shows why fermions (through $m_D$) are essential for nonzero dissipation in the $U(1)$ case.
Our use of the Kubo limit assumes 
$H\ll \Gamma_{\rm th}\sim \alpha\,T$, 
$\omega\sim H \ll m_D$, 
and slow evolution of the background ($|\dot T|/T \ll \Gamma_{\rm th}$, $|\dot Q|/Q \ll \Gamma_{\rm th}$). 
Depending on parameters, either the diffusive regime $\omega \ll m_D^2/\sigma_E$ or the collisionless regime $m_D^2/\sigma_E \ll \omega \ll m_D$ applies. 
These inequalities are satisfied by the benchmarks used in our CMB fits.}.

\section{Warm Inflation}\label{sec:wi}
The evolution of the background axion field is now given in Eq.~\ref{eq:warmbackphi}, which, together with the evolution of the radiation energy density $\rho _R$ and the Friedman equation, completely contain the background information.
\begin{eqnarray}
    \dot{\rho}_R+4H\rho_R=3HQ\dot{\phi}^2\\
    H^2=\frac{1}{3M_p^2}(V(\phi)+\frac{\dot{\phi}^2}{2}+\rho_R)
\end{eqnarray}
These are the most general equations for warm inflation driven by a single field potential $V(\phi)$ for the inflaton $\phi$. Under the assumption of slow-rolling $\phi$ (assumed in adiabatic condition to calculate $\Upsilon $), slowly varying $\rho _R$, and noting that $\rho _{\rm R}, \frac{\dot{\phi } ^2}{2} \ll V(\phi )$ during inflation, the above Eq.s lead to
\begin{eqnarray}\label{eq:SReqs}
\dot{\phi}^2&=&\frac{V_{\phi}^2}{3H^2}\frac{1}{3(1+Q)^2} = \frac{2V\epsilon _V}{3(1+Q)^2},\\
\rho _R &=&\frac{3Q}{4}\dot{\phi}^2= \Mp^2 \frac{V_{\phi}^2}{4V}\frac{Q}{(1+Q)^2} = \frac{V\epsilon _V}{2}\frac{Q}{(1+Q)^2},
\end{eqnarray}
where $\epsilon _V\equiv \frac{\Mp^2}{2}\bigg(\frac{V_{\phi}}{V}\bigg ) ^2$ is the usual slow-roll parameter.

Using $\rho _R = C_R T^4$, where $C_R = \frac{\pi ^2}{30}g_{\star}(T)$, the temperature can be calculated as\footnote{We use $C_R=65.5$ throughout this work, since it corresponds to $g_{\star}=200$.}
\begin{equation}\label{eq:Temp}
T=\bigg(\frac{V\epsilon _V}{2C_R}\bigg)^{1/4}\frac{Q^{1/4}}{(1+Q)^{1/2}}.
\end{equation}

Combining this expression of $T$ with $H^2=V/3\Mp ^2$ in the slow roll limit, 
\begin{eqnarray}\label{eq:Q2}
Q&=&C_Q\lambda^{8/7}\bigg(\frac{m_{\phi}}{f}\bigg)^{2/7}\bigg(\frac{f}{\Mp}\bigg)^{-10/7}\bigg[\frac{\sin ^3(\phi/2f)}{\cos ^2(\phi/2f)}\bigg]^{2/7},\\
&&{\rm where}~~ C_Q=\bigg(\frac{e^3}{576\pi^3}\bigg)^{4/7}\frac{3^{2/7}}{C_R^{3/7}2^{5/7}}.\nonumber
\end{eqnarray}
Due to the weak dependence on the factor in the square bracket, $Q$ remains  nearly constant during inflation. For this reason, in the rest of our analysis, we have considered $Q$ to be a constant. 

\begin{figure}
\includegraphics[width=\linewidth]{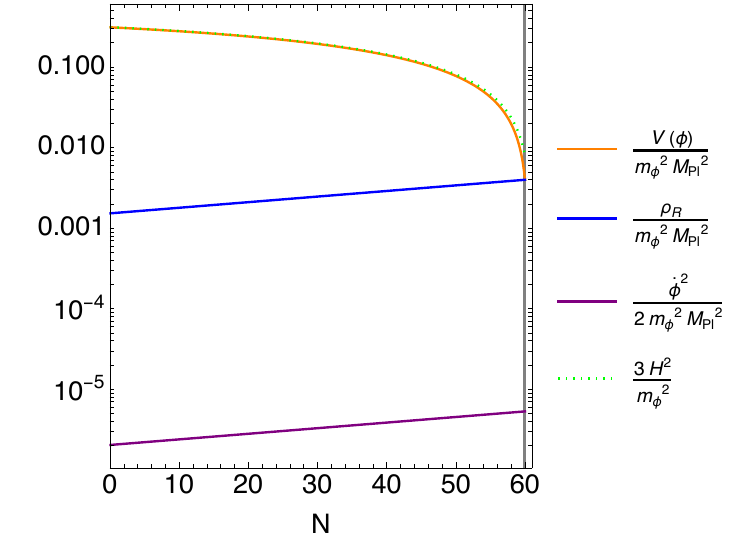}
\caption{Evolution of potential, kinetic and radiation energy densities, normalised by $m_{\phi}^2\Mp^2$ plotted in blue, purple and orange solid lines for the benchmark point $Q=250$, $f=0.5\Mp$, and for $60$ inflationary e-folds. The Hubble parameter is shown with dotted green line and the end of inflation is shown with the vertical gray line.}
\label{fig:energies}
\end{figure}

In this warm setup with a sustaining radiation bath with subdominant energy density, inflation ends with the end of effective slow-roll, i.e., when $\frac{\epsilon}{1+Q}=1$ where $\epsilon=-\frac{\dot{H}}{H^2}\simeq \epsilon_V$. The evolution of energy densities for a benchmark point with constant $Q=200$, $f=0.5\Mp$, $m_{\phi}=2.5\times 10^{-8}\Mp$ and for $55$ inflationary e-folds has been shown in Fig.~\ref{fig:energies}, which shows that inflation indeed ends with the kinetic energy overtaking the potential energ, while the radiation energy density still remains subdominant at the end of inflation. Furthermore, we work in the large dissipation regime, $Q\gg 1$ which is consistent with an ultra-relativistic thermal plasma, and ensures $\rho_R\ll V(\phi)/2$ from Eq.~\eqref{eq:SReqs}. 

The number of e-folds from when a mode $k$ exits the inflationary horizon at e-fold $N(k)$ until the end of inflation at $N_e$ at the field value $\phi_e$ for the periodic potential (\ref{eq:Potential}) is then
\begin{eqnarray}\label{eq:number of efolds}
     N_e-N(k)&= &\int_{N}^{N_e}dN = \gamma^2\ln[\sin(\frac{\phi}{2f})]|_{\phi}^{\phi_e},\\
    {\rm where} \nonumber \\
    \gamma^2&\equiv &\frac{2f^2Q}{\Mp^2}.
\end{eqnarray}
Using the condition $\epsilon(\phi_e)/Q=1$ at the end of inflation, $\sin^2(\phi_e/2f)=\frac{\gamma^2}{1+\gamma^2}$, so that\footnote{Note that, we calculate the number of e-folds forward in time, i.e., $N_e >N(k)$, and therefore $\sin (\phi/2f) < \sin (\phi _e/2f)$, which can be understood remembering that we evolve $\phi /f$ in the first quadrant, and $\phi $ increases as inflation progresses.} using Eq.~\ref{eq:number of efolds},
\begin{equation}\label{eq:phiN}
\sin (\phi/2f) =  \frac{\gamma }{\sqrt{1+\gamma ^2}}\exp \bigg(\frac{N-N_e}{\gamma ^2}\bigg).
\end{equation}
 
The scalar power spectrum in warm inflation has contributions from both vacuum and thermal fluctuations.
\begin{equation}\label{eq:power1}
\mathcal{P}_R \simeq \bigg(\frac{H}{\dot{\phi}}\bigg)^2\frac{H^2}{(2\pi)^2}\bigg[1 + \frac{T}{H}\frac{2\pi Q}{\sqrt{1+4\pi Q/3}}\bigg]\times G(Q).
\end{equation}
The power spectrum can be estimated for the case of strong dissipative regime $Q\gg 1$, where the thermal fluctuations dominate. 
\begin{eqnarray}
    \mathcal{P}_R&=&\frac{Q^{\frac{9}{4}}}{4\sqrt{2\pi^3}C_R^{\frac{1}{4}}\Mp^{\frac{9}{2}}}\frac{V^{\frac{9}{4}}}{V_{,\phi}^{\frac{3}{2}}}G(Q)\label{eq:pow2}\\
    =&P_0&\bigg (\frac{m_{\phi}}{f}\bigg)^{\frac{3}{2}}\frac{\gamma^{\frac{9}{2}}}{g(\gamma)^{\frac{3}{4}}}\bigg(\chi (N,\gamma)-\frac{g(\gamma)}{\chi(N,\gamma)}\bigg)^{\frac{3}{2}}G(Q),\nonumber\\ \label{eq:pow3}\\
    {\rm where} ~ ~ ~ ~ P_0&=&\frac{1}{16\pi^{\frac{3}{2}}C_R^{\frac{1}{4}}}    \nonumber \\
    g(\gamma)&=&\frac{\gamma^2}{1+\gamma^2},~~~{\rm and} ~~\chi(N,\gamma)=\exp\bigg[\frac{N_e-N}{\gamma^2}\bigg]. \nonumber 
\end{eqnarray}
The multiplicative factor $G(Q)$ accounts for the growth or suppression
of inflaton fluctuations due to its direct coupling with radiation~\cite{Montefalcone:2023pvh,Rodrigues:2025neh,Kamali:2023lzq}, which is usually calculated numerically, and then fitted as a function of $Q$. In this work, we estimate it following Eq.(4.8) in~\cite{Montefalcone:2023pvh} as\footnote{This fitting is valid for $Q<10^3$, and therefore valid to use for our work.}
\begin{equation}
G(Q)=1+AQ^{\alpha}+BQ^{\beta},\label{eq:GQ}
\end{equation}
owing to the cubic $T$-dependence found in Eq.~\eqref{eq:Q1}, with $A=3.5\times 10^{-5}, B=3.71\times 10^{-14}, \alpha=6.52$, and $\beta =9.78$.
The second expression for the power spectrum in Eq.~\eqref{eq:pow3} is expressed completely in terms of model parameters $m_{\phi}, f, \gamma (f,Q)$, and number of e-folds using Eq.~\ref{eq:phiN}.

\begin{figure}
\includegraphics[width=\linewidth]{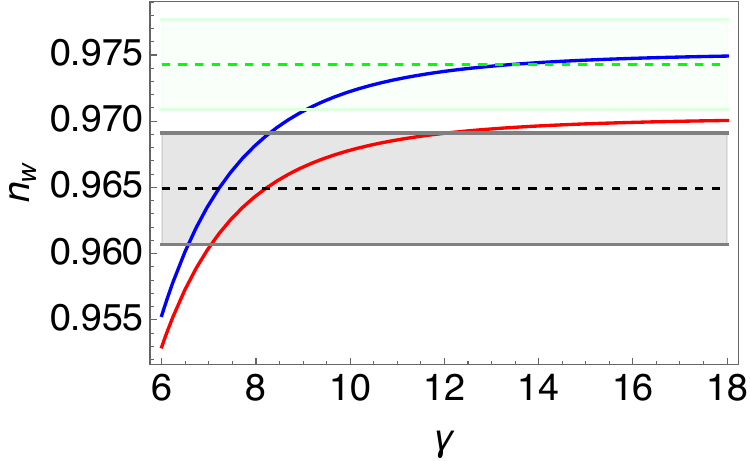}\\
\includegraphics[width=\linewidth]
{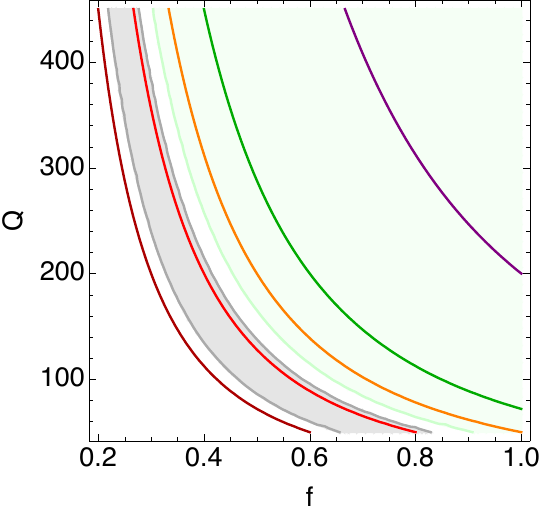}
\caption{\textit{Upper:} Evolution of $n_w$  for $N_{\rm inf}=50$ and $60$ plotted with red and blue lines respectively. The marginalised values of the spectral index from Planck 2018~\cite{Planck:2018jri} $n_s=0.9649\pm0.0042$ and from P-ACT-LB~\cite{AtacamaCosmologyTelescope:2025blo} (CMB lensing from ACT and Planck and BAO data from DESI Y1) $n_s=0.9743\pm0.0034$ are shown with black and green dashed lines respectively. The corresponding $1\sigma$ confidence limits are shown with gray and green shaded regions respectively.\textit{Lower:} Contour plots for $\gamma=6,8,10,12,20$ plotted in solid lines brown, red, orange, dark green, purple. The gray and green shaded regions represent Planck $1\sigma$ and P-ACT-LB $1\sigma$ regions respectively.}
\label{Fig2}
\end{figure}

The CMB pivot scale $k_p=0.05$ Mpc$^{-1}$, crosses the horizon at $N_p$ so that $k_p=a(N_p)H(N_p)$, and therefore, $N_e-N(k) = N_{\rm inf} - \log (k/k_p)$, where $N_{\rm inf}\equiv N_e-N_p$ is the total number of e-folds of inflation between the horizon exit of the pivot scale and end of inflation.
The scalar power spectrum in Eq.~\ref{eq:pow3} can thus be written incorporating the scale dependence as
\begin{eqnarray}\label{eq:pow4}
    \mathcal{P}(k)=&P_0&\bigg (\frac{m_{\phi}}{f}\bigg)^{\frac{3}{2}}\frac{\gamma^{\frac{9}{2}}}{g(\gamma)^{\frac{3}{4}}}\bigg( \bar{\chi}(k,\gamma)-\frac{g(\gamma)}{\bar{\chi}(k,\gamma)}\bigg)^{\frac{3}{2}}G(Q)\nonumber\\
    \\
    \nonumber
  {\rm where} \\
  \bar{\chi}(k,\gamma)&=& \bigg(\frac{k}{k_p}\bigg)^{-\frac{1}{\gamma^2}}\exp[N_{\rm inf}/\gamma^2]\nonumber.
\end{eqnarray}
Using this form of the power spectrum, the amplitude and spectral index at the pivot scale are 
\begin{eqnarray}
A_w &=&P_0\bigg (\frac{m_{\phi}}{f}\bigg)^{\frac{3}{2}}\frac{\gamma^{\frac{9}{2}}}{g(\gamma)^{\frac{3}{4}}}\bigg(\bar{\chi}(k_p,\gamma)-\frac{g(\gamma)}{\bar{\chi}(k_p,\gamma)}\bigg)^{\frac{3}{2}} G(Q)\nonumber\\ \label{eq:Aw}\\
{\rm and} && ~ ~ \nonumber \\
n_w-1&\equiv &\frac{d\log \mathcal{P}(k)}{d\log k}\bigg \vert_{k=k_p}= -\frac{3}{2\gamma ^2} \frac{ \bar{\chi} ^2(k_p,\gamma) + g(\gamma )}{\bar{\chi} ^2(k_p,\gamma) - g(\gamma )}\label{eq:nw}.
\end{eqnarray}
It is interesting to note that the scalar spectral index depends only on the parameter $\gamma (f,Q)$ for a given number of inflationary e-folds. In the upper panel of Fig.~\ref{Fig2}, $n_w$ is potted with $\gamma$ for two values of inflationary number of e-folds.
In particular, an asymptotic behavior of $n_w$ in the limit of large $\gamma (f,Q)$ is noted
\begin{equation}
n_w\vert_{\gamma \gg 1}\simeq \frac{2(N_{\rm inf}-1)}{2N_{\rm inf}+1},
\end{equation}
so that $n_w^{\rm max}\vert _{N_{\rm inf}=60}\simeq 0.9752$. This maximum value is at $>3\sigma $ confidence limit (C.L.) of Planck data combination (see caption of Fig.~\ref{Fig2}). The clear dependence of $n_w(f,Q)$ is shown as contour plot in the lower panel of Fig.~\ref{Fig2}, which shows that for sub-Planckian $f$, large $Q$ values are allowed to lead to a CMB consistent scalar spectral index with Planck 2018 data. 
This figure shows that CMB constraints with Planck data are expected to be satisfied for $\gamma^2\sim \mathcal{O}(100)$, which can be obtained for sub-Planckian $f$ only in the strong dissipative regime. \footnote{The contour plot agrees with this conclusion for the Planck data. However, for the recent P-ACT-LB data, the contours point to arbitrarily large $f$ and $Q$, which may make it difficult to probe this model including the P-ACT-LB likelihood. We leave it for the future, pending resolution for the tendency of the BAO data to push the scalar spectral index to higher values}. In the contour plot, the largest $\gamma (f=1,Q=450)$ leads to $n_w\simeq 0.9752$, and therefore, the theoretical limit to reach $n_w^{\rm max}$ has been pushed here.

With these equations and estimates,
we are now ready to enter our model in CAMB with a viable prior range and test it using CMB data. 
 
\section{Analysis}\label{sec:analysis}
We analyse our model by implementing the power spectrum in~Eq.\ref{eq:pow4} in the cosmological Boltzmann code \texttt{CAMB}~\cite{Lewis:1999bs,Howlett:2012mh}. We fix the value of $N_{\rm inf}$ and find the best fit values of the parameters $f,Q$ and $m_{\phi}/f$.
For our cosmological analysis, we constrain the inflation parameters alongside the late time cosmological parameters \mbox{$\{\Omega_{\rm b}h^2,\Omega_{\rm c}h^2,\theta , \tau \}$} using the following datasets:
\begin{itemize}
    \item \textit{Planck} 2018 TT + TE +EE  with low$\ell$ and low $E$ likelihood~\cite{Aghanim:2019ame};
    \item  \textit{Planck} 2018 lensing likelihood~\cite{Aghanim:2018oex};
    \item  SDSS BAO likelihood~\cite{BOSS:2016wmc} .
\end{itemize}
The likelihoods are sampled using the \texttt{MCMC} sampler~\cite{Lewis:2002ah,Lewis:2013hha}, through its interface with \texttt{CosmoMC}. We sample the likelihoods until we reach a value $R-1=0.05$ for the Gelman-Rubin diagnostic. The resulting chains are analysed and plotted with~\texttt{GetDist}~\cite{Lewis:2019xzd}. At the end of the sampling, we also run a
minimizer to find the best-fit point and the corresponding $\chi^2$ values. 

From Eq.~\ref{eq:pow4}, it is evident that the power spectrum without $G(Q)$ depends only on the following combinations of the model parameters:$\frac{m_{\phi}}{f}$ and $f^2 Q$. However, including $G(Q)$ is necessary here due to large $Q$ (see Eq.~\eqref{eq:GQ}). Therefore, using the full expression from Eq.~\ref{eq:pow4}, it is possible to constrain $f,Q$ and $m_{\phi}/f$ separately.
Noting that the amplitude of scalar power spectrum depends on $m_{\phi}/f$ and has stringent constraints from CMB, we choose to vary ($f,10^{-3}Q,\ln(10^{20}m_\phi/f)$) in CAMB.
 \begin{figure}
\includegraphics[width=0.8\linewidth]{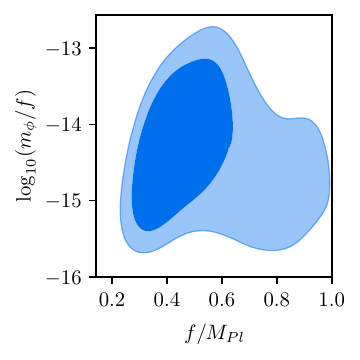}\\
\includegraphics[width=0.8\linewidth]{
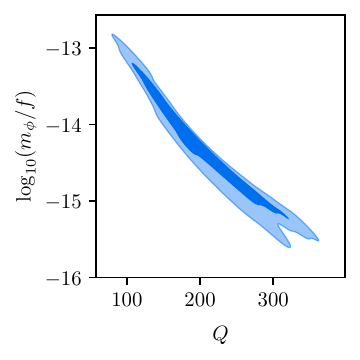}\\
\caption{2D posterior distribution of the derived model parameters. Dark and light blue contours represent $1\sigma$ and $2\sigma$ confidence limits (C.L.) respectively.}
\label{fig:derivedparams}
\end{figure}
\begin{table}[ht]
\centering
\caption{Marginalised mean values and best-fit values of the key model parameters derived from the analysis}
\begin{tabular}{|c|c|c|c|}
\hline
  &$\log_{10}(m_{\phi}/f)$ & $f$& $Q$ \\
\hline 
\hline
Mean& $-14.34$ & $0.49$ &$208.43$ \\
\hline
1$\sigma $&$0.59$ & $0.16$ &$59.47$  \\
\hline
Best-fit & $-14.29$ & $0.41$ & $201.64$ \\
\hline
\end{tabular}
\label{tab1}
\end{table}

In Fig.~\ref{fig:derivedparams} we plot the two-dimensional posterior probability distribution for the derived parameters  $\frac{m_{\phi}}{f}$ with $f$ and $Q$ for $N_{\rm inf}=60$. Marginalised over all the other parameters, their mean values corresponding $1\sigma$ C.L., and best-fit values are quoted in Table~\ref{tab1}. Figure (\ref{fig:triangle_plot}) and Table~\ref{table:allparams} in the Appendix~\ref{APP2} present the posterior probability distribution of all 7 cosmological parameters and their mean values  with $1\sigma$ confidence limits (along with best-fit values) respectively. 

\begin{figure}[h]
\includegraphics[width=0.8\linewidth]{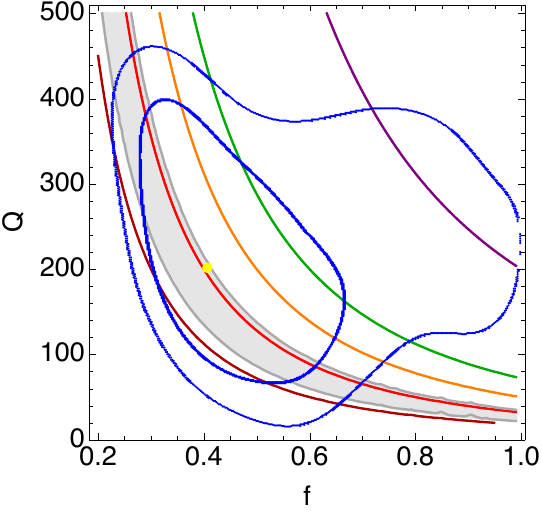}
\caption{2D posterior distribution for $f-Q$ with  $1\sigma$ and $2\sigma$ C.L. (in blue dots) obtained from MCMC analysis, embossed on the $f-Q$ contour plots. The best-fit ${f,Q}$ point is plotted in yellow dot. See the caption of Fig.~\ref{Fig2} (lower panel) for the plot legends.}
\label{fQcontourcombo}
\end{figure}

The top panel of Fig~\ref{fig:derivedparams} and Table~\ref{tab1} shows that $f\geq\Mp$ is excluded by $>3\sigma$ C.L.\footnote{One can ask the prior dependence for this result, since we used a flat prior $0.1\leq f/\Mp\leq 1$. However, this choice of priors in justified for large $Q$ (prior $80\leq Q \leq 450$), as discussed with respect to Fig.~\ref{Fig2}.}. The spectral index obtained with the mean values are $n_w(f_{\rm mean},Q_{\rm mean})\simeq 0.9722$, whereas with the best-fit values, $n_w(f_{\rm best-fit},Q_{\rm best-fit})\simeq 0.9687$. The bottom panel of Fig.~\ref{fig:derivedparams} also dictates a tight upper bound on $Q$. This is expected, since large $Q$ contributes increasingly to $A_w$ through $G(Q)$, see Eq.~\eqref{eq:Aw}. The (negative) correlation in the bottom panel of Fig.~\ref{fig:derivedparams} can be understood by the leading contribution $(m_{\phi}/f)^{3/2}\times G(Q)$ in Eq.~\eqref{eq:Aw}. In fact, the upper bound on $Q$, combined with our understanding of the contour plot (lower panel of Fig.~\ref{Fig2}), concludes a lower limit of $f/\Mp \geq 0.17$ at $2\sigma$. This is an interesting phenomenological constraint to the limit in reaching for sub-Planckian $f$ when confronted with the latest observations of CMB. The joint posterior probability distribution of $f$ and $Q$ are plotted with the $f-Q$ contours for $\gamma$ in Fig.~\ref{fQcontourcombo}.

\section{Conclusions and Discussions}\label{sec:condis}
In this work, we explored the constraints from CMB observations on the axion $U(1)$ inflation realized in the warm paradigm. In Sec.~\ref{sec:themodel}, starting from the equation of motion~\eqref{eq:Backgroung E.O.M}, we have assumed the source due to the large fluctuations in the gauge sector to be thermal. Thus, the source term leads to the dissipation from the inflaton field to the electromagnetic sector, containing gauge fields and fermions, with a dissipation coefficient $\Upsilon$ which is estimated in Eq.~\eqref{eq:upsilonexpression} using linear response method. The result for the dissipation coefficient in warm axion $U(1)$ inflation is a novelty of our work. 

The form of the dissipation coefficient $\Upsilon$ is illuminating further than the cubic dependence on temperature found in our scenario. The fermion mass doesn't appear explicitly in this expression~\eqref{eq:upsilonexpression} since the plasma is ultra-relativistic. However, the contribution of the fermion is important, since it induces thermal mass to the gauge field, which contributes to the dissipation as $\Upsilon \propto m_D^3$. This dependence on the thermal mass clearly indicates that for a simpler scenario, with no fermions present (e.g. for a dark $U(1)$ sector), and therefore $m_D=0$, the dissipation coefficient is vanishing. In other words, fermions play a crucial role in our setup by inducing a thermal mass of the photons which directly leads to a non-zero thermal dissipation.

Furthermore, the Chern-Simons coupling is dependent on the fermionic charge $e$, such that $\lambda =q\alpha$, where $q$ is a constant that varies from model to
model~\cite{Dimastrogiovanni:2023juq}, and $\alpha=e^2/(4\pi)$.
Therefore, we obtain 
$\Upsilon\propto e^{7}T^3/f^2$. In the strict $\omega\!\to\!0$ (diffusive) limit the normalization acquires a mild $\ln(1/e)$ from $\sigma_E$, while in the collisionless window $(m_D^2/\sigma_E)\ll \omega \ll m_D$ the same $T^3$ scaling holds without the log. 
For comparison, non-Abelian sphaleron heating yields $\Upsilon\propto g^{10}T^3/f^2$ ~\cite{McLerran:1990de,Laine:2021ego,Papageorgiou:2022prc}. This distinct dependence $\Upsilon(e)$ is characteristic of a warm axion-inflation coupled to Abelian U(1) gauge fields.


In Sec.~\ref{sec:wi}, the dynamics of warm inflation has been explored in the context of our model. With the justified assumption of a constant temperature \footnote{For our best-fit region, $\Gamma_{\rm th}\sim \alpha T$ satisfies $\Gamma_{\rm th}/H \gtrsim \mathcal{O}(10^2)$, ensuring adiabaticity of the bath and validating the use of the Kubo limit at $\omega\sim H$.}, hence constant $Q$, as discussed with Eq.~\eqref{eq:Q2}, we proceed to find the dependence of the inflationary power spectrum on the model parameters $m_{\phi}/f$, $f/\Mp$, and $Q$ (Eq.~\eqref{eq:pow4}). In the same section, the effect of the model parameters on the pivot amplitude and spectral index of the scalar power spectrum have also been discussed in detail. Using this expression in the \texttt{CAMB} code, we then constrain the full model with the 3 inflationary parameters and 4 standard parameters of $\Lambda$CDM model with \texttt{CosmoMC}, using cosmological data as listed in section~\ref{sec:analysis}. Our result shows the preferred posterior distribution for different combinations of parameters $m_{\phi}/f$, $f/\Mp$, and $Q$. The conclusions on the model parameters drawn from this analysis are discussed at the end of Sec.~\ref{sec:analysis}.

\begin{figure}[h]
\includegraphics[width=0.8\linewidth]{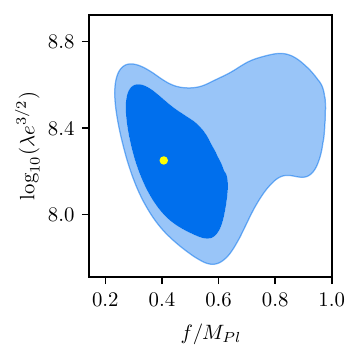}
\caption{2D posterior distribution for $\lambda e^{3/2}-f$ with  $1\sigma$ and $2\sigma$ C.L. obtained from MCMC analysis, embossed on the $f-Q$ contour plots. The best-fit ${\lambda e^{3/2},f}$ point is plotted as an yellow dot.}
\label{derived_lam_f}
\end{figure}

It is important to point out here that while deriving the dissipation coefficient using linear response theory, we have not used the linear solution for the gauge modes $A_\mu$. The only assumption that is necessary for our derivation is the presence of large homogeneous back-reaction\footnote{To clarify, homogeneous back-reaction in the context of axion inflation refers to the gauge fields coupled to a homogeneous inflaton(axion)~\cite{Barbon:2025wjl,Figueroa:2023oxc}.}, so that the inflaton still follows slow-roll dynamics. The hierarchy of the different terms present in the equation of motion is such that the back-reaction, i.e., thermal friction term here containing $\Upsilon$ (or $Q$) in Eq.~\eqref{eq:warmbackphi} is the dominating over the others, while preserving slow-roll of the axion (see Fig.~\ref{fig:backreacterms} at the end of Appendix~\ref{APP1}).We can also estimate the large value of the coupling constant $\lambda$ required in this case using the constraints on $m_{\phi}/f$, $f/\Mp$, and $Q$, and Eq.~\eqref{eq:Q1}. The 2D posterior probability distribution for the derived $\lambda e^{3/2}$ and $f$ is plotted in Fig.~\ref{derived_lam_f}.

It is, nevertheless, interesting to investigate the status of the `particle production' parameter $\xi \equiv \frac{\lambda \dot{\phi}}{2fH}$, that signifies the instabilities of the gauge fields. In cold axion inflation models, $\xi$ is the parameter that leads to the growth of one helicity of the gauge fields in the linear solution, $A_{+} \sim \frac{1}{\sqrt{2k}}e^{\pi \xi /2}W_{-i\xi,1/2}(2ik\tau)$~\cite{Anber:2009ua,Barnaby:2011vw}, where $W_{a,b}(x)$ is a Whittaker function. This solution is valid in the limit, $\xi \lesssim 4.7$,~\cite{Peloso:2016gqs} for the back-reaction to be small, whereas additional limits, such as from non-Gaussianity~\cite{Barnaby:2012xt,Ozsoy:2021onx} and perturbativity~\cite{Peloso:2016gqs} are of comparable order.
Rewriting the first slow roll equation~\eqref{eq:SReqs} in the limit of large $Q$ as
$
\frac{\dot{\phi}}{H}\simeq \frac{\Mp^2}{fQ}\tan (\phi/2f)$
leads to the following parametric dependence of $\xi$
\begin{equation}
\xi = \frac{\lambda}{\gamma ^2}\tan (\phi/2f).\label{eq:xi2}
\end{equation} 
Using Eq.~\eqref{eq:xi2} and~\eqref{eq:phiN} with the best-fit values of $f$, $Q$ and the derived parameter $\lambda$ (assuming $e=1$), $\xi _{\rm best-fit}^{\rm pivot}\approx 10^6$. It is evident, therefore, that demanding the thermal back-reaction to be large enough to drive a warm $U(1)$-axion inflation in the strong dissipation regime, we are beyond the limits on $\xi$ deduced using the cold linear solution.


Ref.~\cite{Ferreira:2017wlx} has discussed the bounds on $\xi$ in the $\xi$-$f/\lambda H$ space for the onset of a thermal (warm) paradigm during axion-$U(1)$ inflation, including the backreaction and perturbativity bounds discussed above, charting the parameter space where it is possible to thermalize (with and without fermionic interactions)\footnote{In~\cite{Ferreira:2017wlx} $\lambda =1$, and therefore in our context the y-axis of their Figure 4 should be $f/\lambda H$ instead of $f/H$.}. However, they have explored these limits assuming a cold (non-thermal) initial condition, within the realm of the linear solution, to reach conditions on the parameter space for gradually attaining thermalization during inflation, whereas in this work we have assumed thermalization as an initial condition itself. Other than this key difference, let us also mention that the primordial power spectrum in~\cite{Ferreira:2017wlx} is from quantum vacuum flucutations of the axion field, whereas our power spectrum is a completely thermal one. With the best-fit values of the model parameters and derived parameters evaluated in this section and the previous one, we belong to far top right corner of their parameter space (see Figure 4 in~\cite{Ferreira:2017wlx}), which is expected as we do not follow the linear solution in the weak backreaction regime. The limit $\xi \gtrsim 2.9$ found in their analysis to satisfy SM thermalization is also relevant here, indicating that we possibly also have a thermalized axion in this case. 


For the best-fit value of the model parameters, we find $m_{\phi}\simeq 5$ TeV, and the following energy hierarchy holds: $H\lesssim m_{\phi}\ll \Lambda \ll f$, where $\Lambda \equiv 2\xi H=\lambda \dot{\phi}/f$ is the effective energy scale in presence of a Chern-Simons coupling. The inflaton mass scale favored by the phenomenologically viable region of parameter space,
$m_\phi = \mathcal{O}(\mathrm{TeV})$,
lies close to the electroweak scale. This makes the presence of Standard Model charged particles in the thermal bath plausible during inflation and avoids introducing hierarchically separated energy scales. Moreover, the approximate shift symmetry associated with the axion-like nature of $\phi$ protects the inflaton potential against large perturbative thermal mass corrections, ensuring the technical naturalness of the scenario.

The stability of the charged fermions is not fixed by the inflationary effective Lagrangian alone. In the baseline interpretation used in this paper, $\psi$ denotes a vectorlike charged plasma degree of freedom during inflation and need not be stable at late times; portal interactions or additional lighter states in a UV completion can make it decay before it contributes to late-time relic abundances. If instead an exact fermion-number symmetry or a $\mathbb{Z}_2$ symmetry makes $\psi$ stable, then it becomes a dark-sector relic and its abundance has to be included. As a simple estimate, for a stable Dirac fermion annihilating into two Abelian gauge bosons,
\begin{equation}
\langle \sigma v\rangle_{\psi\bar\psi\rightarrow AA}\simeq \frac{\pi\alpha_D^2}{m_f^2},\qquad \alpha_D\equiv \frac{e^2}{4\pi},
\end{equation}
and the thermal relic abundance scales as
\begin{equation}
\Omega_\psi h^2\simeq 0.12\left(\frac{2.2\times10^{-26}\,\mathrm{cm^3s^{-1}}}{\langle\sigma v\rangle}\right).
\end{equation}
Consequently, a stable $\psi$ would impose extra constraints on $m_f$, $e$, the post-inflation temperature of the $U(1)$ sector, and possible kinetic-mixing or portal interactions. If the $U(1)$ is identified with an ordinary electromagnetic or hypercharge direction, a stable charged relic would be phenomenologically unacceptable unless it decays efficiently; if it is a hidden $U(1)$, $\psi$ may instead be a viable dark-matter candidate in a restricted parameter region. A dedicated relic-density and dark-radiation analysis depends on the UV completion and lies beyond the CMB-focused inflationary constraints presented here; we regard it as a natural follow-up project.

For non-Abelian gauge fields, recently~\cite{Kamali:2024qme} have explored a the chromonatural inflation model assuming a priori thermalization, whereas ~\cite{Bhattacharya:2025guc} have explored the limits in the parameter space where such an assumption is justified. For both Abelian and non-Abelian gauge fields present during axion inflation, the exact evolution from a cold to warm scenario either before or after the horizon exit of the pivot scale during inflation is however an important piece of the puzzle that is yet to be explored. The inherent discrepancy between the dynamics of a fully cold and fully warm axion inflation (with Abelian/non-Abelian gauge fields) calls for a bridge between the two paradigms, distinctly showing in the parameter space the continuous transition between them. This is a difficult task, since moving away from the cold linear solution is extremely difficult to handle, even with numerical/lattice techniques (see ref.s~\cite{Caravano:2021bfn,Caravano:2022epk,Figueroa:2023oxc,Figueroa:2024rkr,Caravano:2024xsb,Sharma:2024nfu}). As a distinct phenomenological signature for a warm $U(1)$-axion inflation explored here, the non-Gaussianities are expected to have distinct shapes~\cite{Bastero-Gil:2014raa,Mirbabayi:2022cbt}, which we hope to explore in the future. 

\section*{Acknowledgement}
SB acknowledges the “Consolidaci\'{o}n Investigadora” grant CNS2022-135590 and the support from the Spanish Research Agency (Agencia Estatal de Investigaci\'{o}n)
through the Grant IFT Centro de Excelencia Severo Ochoa No CEX2020-001007-S, funded by
MCIN/AEI/10.13039/501100011033. AA, SB and VK  thank R. Brandenberger, M. Fasiello,  and A. Papageorgiou for illuminating discussions and helpful comments. SB acknowledges helpful comments from M. Bastero-Gil and G. Montefalcone on the first version of the draft.
\bibliography{ref}
\appendix
\section{Estimation of the dissipation coefficient }
\label{APP1}
The dissipation coefficient $\Upsilon$ plays a critical role in understanding the dynamics of a scalar field $\phi$ interacting with other fields $X$. If the potential is given as:
\begin{equation}
\nonumber
    V(\phi, X) = V(\phi) + V_{\text{int}}(\phi, X), 
\end{equation}
where the interaction term is:
\begin{equation}
\nonumber
    V_{\text{int}}(\phi, X) = f(\phi) g(X). 
\end{equation}
then the effective equation of motion for $\phi$ becomes:
\begin{eqnarray}
\nonumber
        \frac{\partial}{\partial \mu} \frac{\partial \mathcal{L}_{\text{eff}}}{\partial (\partial_\mu \phi)} 
    - \frac{\partial \mathcal{L}_{\text{eff}}}{\partial \phi} 
    - i \frac{\partial f(\phi)}{\partial \phi} \int d^4x' \theta(t - t') \nonumber 
    \\
    \times \big[f(\phi(x')) - f(\phi(x))\big] \langle [g(X(x)), g(X(x'))] \rangle = 0.\nonumber 
\end{eqnarray}
With the adiabatic approximation $\frac{\dot{\phi}}{\phi} \ll \tau^{-1}$, the field's evolution is slow compared to the system's response timescale $\tau$ \cite{Bastero-Gil:2010dgy}.

Expanding $f(\phi)$ to estimate the causal effect at a time $t'$ around time $t$ using a Taylor series
\begin{equation}
\nonumber
    f(\phi(t')) - f(\phi(t)) \approx (t' - t) \dot{\phi}(t) \frac{\partial f(\phi)}{\partial \phi}, 
\end{equation}
and incorporating dissipative terms, leads to the effective equation of motion Eq.~\eqref{eq:warmbackphi}. The dissipation coefficient $\Upsilon$ is defined as:
\begin{equation}
\nonumber
    \Upsilon = \int d^4x' \Sigma_R(x, x') (t' - t). 
\end{equation}
The retarded correlation function $\Sigma_R(x, x')$ is given by:
\begin{equation}
\nonumber
    \Sigma_R(x, x') = -i \left[\frac{\partial f(\phi)}{\partial \phi}\right]^2 \theta(t - t') \langle [g(X(x)), g(X(x'))] \rangle. 
\end{equation}
These equations describe how energy dissipates from the scalar field $\phi$ into coupled fields $X$, significantly influencing the field's return to equilibrium.

The retarded Green's function for an operator \( O(t, \mathbf{x}) \) is defined as:
\begin{equation}\nonumber
\Sigma_R(t-t', \mathbf{x}-\mathbf{x}') = i \Theta(t-t') \langle [O(t, \mathbf{x}), O(t', \mathbf{x}')] \rangle,
\end{equation}
where $\Theta(t-t')$ is the Heaviside step function, ensuring causality. This definition captures the causal response of the system to a perturbation at \( t' \).

To study the Green's function in momentum space, we take the Fourier transform:
\begin{equation}\nonumber
\Sigma_R(t-t', \mathbf{x}-\mathbf{x}') = \int \frac{d\omega}{2\pi} e^{-i\omega(t-t')} \int \frac{d^3k}{(2\pi)^3} e^{i \mathbf{k} \cdot (\mathbf{x}-\mathbf{x}')} \tilde{G}^R(\omega, \mathbf{k}).
\end{equation}

The thermal average of $ O(t, \mathbf{x}) O(t', \mathbf{x}') $ can be expanded using the eigenstates $|n\rangle $ and $|m\rangle $ of the unperturbed Hamiltonian:
\begin{equation}\nonumber
\langle O(t, \mathbf{x}) O(t', \mathbf{x}') \rangle = \sum_{n,m} \langle n| O(t, \mathbf{x}) |m \rangle \langle m | O(t', \mathbf{x}') |n \rangle e^{-\beta E_n}.
\end{equation}

Substituting $O(t, \mathbf{x}) = e^{iHt - i \mathbf{P} \cdot \mathbf{x}} O(0, 0) e^{-iHt + i \mathbf{P} \cdot \mathbf{x}}$, we can write:
\begin{eqnarray}  
\nonumber
\langle O(t, \mathbf{x}) O(t', \mathbf{x}') \rangle = \sum_{n,m} e^{-i (E_m - E_n)(t-t') + i (\mathbf{p}_m - \mathbf{p}_n) \cdot (\mathbf{x} - \mathbf{x}')}\\
\nonumber
\times e^{-\beta E_n} |\langle n | O(0, 0) | m \rangle|^2.
\end{eqnarray}

The commutator $ \langle [O(t, \mathbf{x}), O(t', \mathbf{x}')] \rangle$ therefore becomes:
\begin{eqnarray}
\nonumber
   \langle [O(t, \mathbf{x}), O(t', \mathbf{x}')] \rangle = \sum_{n,m} e^{-i (E_m - E_n)(t-t') + i (\mathbf{p}_m - \mathbf{p}_n) \cdot (\mathbf{x} - \mathbf{x}')} \\
\nonumber
\times e^{-\beta E_n} (1 - e^{-\beta \omega}) |\langle n | O(0, 0) | m \rangle|^2,
 \end{eqnarray}
where $\omega = E_m - E_n $.

Taking the Fourier transform of the commutator gives:

\begin{eqnarray}
\rho_0(\omega, \mathbf{k}) &\equiv& \text{FT}(\langle [O(t, \mathbf{x}), O(t', \mathbf{x}')] \rangle) \nonumber \\
&=& (1 - e^{-\beta \omega}) \text{FT}(\langle O(t, \mathbf{x}) O(t', \mathbf{x}') \rangle),\nonumber
\end{eqnarray}
where the Bose-Einstein distribution $ n_B(\omega) = 1/(e^{\beta \omega} - 1) $ relates the occupation of states to temperature.

$\rho_0(\omega, \mathbf{k})$ is the spectral density function for the operator $O$. This function encapsulates the thermal behavior of the system and is essential for determining the plasma's retarded Green’s function. The spectral density function for our case is then evaluated with the prescription of~\cite{Hassan:2024bvb}, leading to the expression in Eq.~\eqref{eq:Greenstemp}.
\\
\begin{figure}[h!]
\includegraphics[scale=0.75]{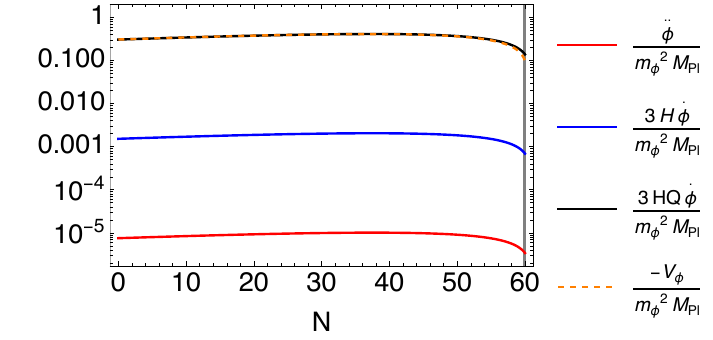}
\caption{Hierarchy of different terms in Eq.~\eqref{eq:warmbackphi} to show that thermal friction dominates, while preserving slow-roll.}
\label{fig:backreacterms}
\end{figure}

\section{Vanishing of the dissipation coefficient without charged fermions}
\label{APPfermionZero}
This appendix records the explicit small-frequency check to reach Eq.~\eqref{eq:upsilonzero_main}. The dissipative term in the warm equation of motion is controlled by the zero-momentum Kubo limit
\begin{equation}
\Upsilon=\frac{\lambda^2}{(4\pi f)^2}\lim_{\omega\rightarrow0}\frac{1}{\omega}\rho^{\rm ret}_{E\!\cdot\!B\,E\!\cdot\!B}(\omega,\mathbf{k}=0),
\end{equation}
so a nonzero warm friction coefficient requires a spectral density proportional to $\omega$ at small frequency. In an Abelian plasma this linear term is generated by the HTL response of charged matter. For $N_f$ relativistic Dirac fermions,
\begin{equation}
 m_D^2=\frac{e^2N_fT^2}{3},
\end{equation}
whereas in the absence of charged fermions $N_f=0$ and the photon polarization tensor has no Debye-screened or Landau-damping part.

This can be seen directly by setting $m_D=0$ in the spectral expression used in Eq.~\eqref{eq:Greenstemp}. For positive $\omega$, the leading free-photon contribution becomes
\begin{eqnarray}
\tilde\Sigma^{(0)}_{\rho}(\omega,\mathbf{0})&=&-i\frac{\lambda^2}{768\pi^3f^2}\omega^4(1-e^{-\beta\omega})\nonumber\\
&&\times\left[\frac{1}{e^{\beta\omega/2}-1}+\frac{1}{(e^{\beta\omega/2}-1)^2}\right].
\label{eq:sigma_nofermion}
\end{eqnarray}
With $x=\beta\omega\ll1$,
\begin{equation}
(1-e^{-x})\left[\frac{1}{e^{x/2}-1}+\frac{1}{(e^{x/2}-1)^2}\right]
=\frac{4}{x}+\mathcal{O}(1),
\end{equation}
and hence
\begin{equation}
\tilde\Sigma^{(0)}_{\rho}(\omega,\mathbf{0})
=-i\frac{\lambda^2}{192\pi^3f^2}T\omega^3+\mathcal{O}(\omega^4).
\end{equation}
Using the definition of $\Upsilon$ then gives
\begin{equation}
\Upsilon_{N_f=0}=i\lim_{\omega\rightarrow0}\frac{\partial\tilde\Sigma^{(0)}_{\rho}(\omega,\mathbf{0})}{\partial\omega}=0.
\end{equation}
Thus the charged fermions are essential not because they make the Chern-Simons interaction gauge invariant, but because they generate the finite-temperature Abelian plasma response. Without them there is no Debye mass, no conductivity, no Landau damping branch cut in the $U(1)$ propagator, and therefore no local warm-friction coefficient of the form used in Eq.~\eqref{eq:warmbackphi}.

\section{Error in assuming $Q$ to be constant}
\label{APP3}
Here we schematically evaluate the error propagating in the system due to the assumption of a constant $Q$ instead of the full form of Eq.~\eqref{eq:Q2}. We can write
\begin{eqnarray}
\frac{d\log Q}{dN}&=&\frac{6}{7}\frac{d}{dN}\log(\sin (\frac{\phi}{2f})) - \frac{4}{7}\frac{d}{dN}\log(\cos (\frac{\phi}{2f}))\nonumber \\
&=&\bigg[\frac{3}{7}\frac{1}{\tan(\frac{\phi}{2f})}+\frac{2}{7}\tan(\frac{\phi}{2f})\bigg]\frac{\dot{\phi}}{fH}\nonumber\\
&\lesssim & \mathcal{O}(\frac{\phi}{f}).
\end{eqnarray}
Therefore, the logarithmic evolution of $Q$ is proportional to the normalised effective energy scale, $\Lambda /H$. However, the last expression in the $\phi/f$-dependence in the square bracket in the penultimate expression shows that caution should be maintained for the cases with large field excursion. In this work, we explore the constant-$Q$ realm, assuming this effect to be small enough. We hope to attempt more thorough analysis in future.

\newpage
\onecolumngrid
\section{Detailed results for posterior distributions}
\label{APP2}
\begin{figure}[h!]
\includegraphics[scale=0.35]{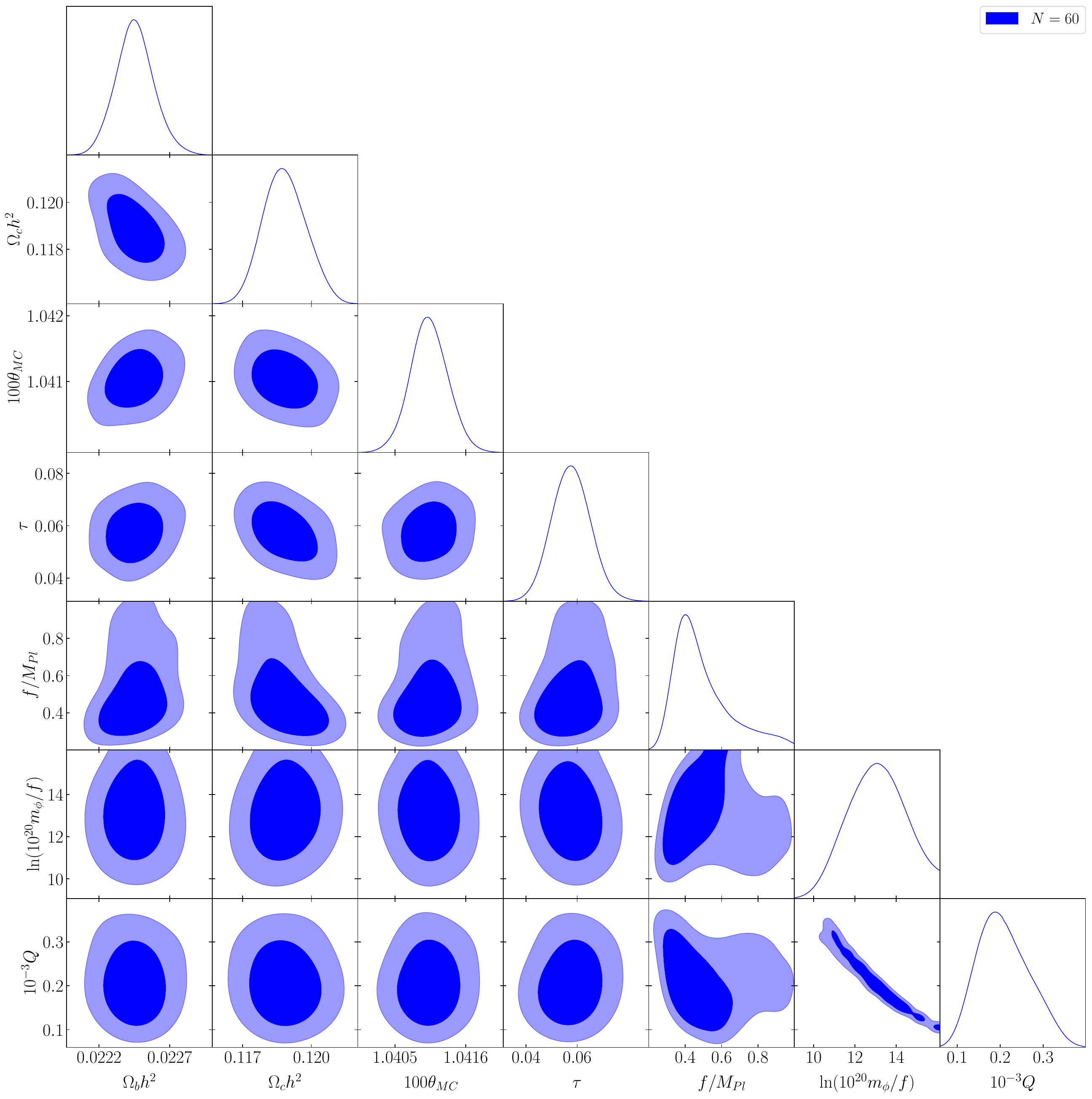}
\caption{Posterior distributions for all parameters.}
\label{fig:triangle_plot}
\end{figure}
\begin{table}[h!]
\begin{tabular} {| l| c |}
\hline
Parameter & $N_{\rm inf}= 60$\\
\hline
\hline
{\boldmath$\Omega_b h^2   $} & $0.02245\pm 0.00014    ~(0.02238)    $ \\

{\boldmath$\Omega_c h^2   $} & $0.11883^{+ 0.00088}_{-0.00100}    ~(0.11927)    $\\

{\boldmath$100\theta_{MC} $} & $1.04103\pm 0.00029   ~(1.04113)     $\\

{\boldmath$\tau           $} & $0.0576 \pm 0.0074~(0.0604)$\\

{\boldmath${\rm{ln}}(10^{20} m_{\phi}/f)$} & $13.03\pm 1.36~(13.14) $\\

{\boldmath$f/\Mp            $} & $0.494^{+0.07}_{-0.19}~(0.406)$\\

{\boldmath$10^{-3}Q            $} & $0.208^{+0.051}_{-0.071}~(0.202)$\\
\hline
\end{tabular}
\caption{Mean values with $68\%$ limits of all the parameters. The corresponding best-fit values are given in brackets. }
\label{table:allparams}
\end{table}

\end{document}